# Evolution of the bilayer ν = 1 quantum Hall state under charge imbalance

W.R. Clarke[1†], A.P. Micolich[1], A.R. Hamilton[1], M.Y. Simmons[1], C.B. Hanna[2], J.R. Rodriguez[2], M. Pepper[3] and D.A. Ritchie[3]

[1]*School of Physics, University of New South Wales, Sydney NSW 2052, Australia.*
[2]*Department of Physics, Boise State University, Boise ID 83725-1570.*
[3]*Cavendish Laboratory, University of Cambridge, Cambridge, CB3 0HE, U.K.*

We use high-mobility bilayer hole systems with negligible tunneling to examine how the bilayer ν = 1 quantum Hall state evolves as charge is transferred from one layer to the other at constant total density. We map bilayer ν = 1 state stability versus imbalance for five total densities spanning the range from strongly interlayer coherent to incoherent. We observe competition between single-layer correlations and interlayer coherence. Most significantly, we find that bilayer systems that are incoherent at balance can develop spontaneous interlayer coherence with imbalance, in agreement with recent theoretical predictions.

The integer [1] and fractional [2] quantum Hall (QH) effects highlight the role of quantization and interactions in two-dimensional (2D) electron systems. Bilayer systems consist of two parallel, closely spaced 2D electron layers separated by an insulating barrier, with a total carrier density $n_T = n_t + n_b$, where $n_t$ and $n_b$ are the carrier densities of the top and bottom layers, respectively. Bilayer QH states, analogous to those in single 2D layers, are observed at integer and fractional total filling factors $\nu = hn_T/eB$ (= $\nu_t + \nu_b$), where $B$ is the applied perpendicular magnetic field [3,4]. These bilayer QH states can be attributed to one or more of three mechanisms: (i) the addition of two independent single-layer QH states at even integer bilayer filling factors [5]; (ii) the symmetric-antisymmetric gap $\Delta_{SAS}$ established by interlayer tunneling at odd integer bilayer filling factors [5]; and (iii) spontaneous interlayer coherence (SILC), which produces remarkable bilayer QH states with no counterpart in single-component systems [3-5].

The bilayer ν = 1 SILC QH state, hereafter referred to as the SILC state, results from correlations between electrons in different layers that unify the two layers into a single system when the layer separation is sufficiently small [4]. The relative strength of the interlayer and intralayer interactions governs the existence of the SILC state, and is parameterized by the distance ratio $d/l_B$, where $d$ is the layer separation and $l_B = (\hbar/eB)^{1/2} = (2\pi n_T)^{-1/2}$ is the magnetic length at ν = 1. It is possible to alter $n_T$, and hence $d/l_B$, electrostatically – a technique used in earlier studies of balanced ($n_t = n_b$) bilayer electron systems to show that the SILC state disappears when interlayer interactions are weakened sufficiently ($d/l_B > \sim 1.8$) [6]. Significant attention is currently focused on understanding how the SILC state evolves as the electron densities in the two layers are unbalanced at constant $n_T$ (i.e., constant $d/l_B$) by transferring charge from one layer to the other – a process known as charge imbalance. A key question is: for $d/l_B > 1.8$, where the SILC state is *not* observed at equal layer densities, can charge imbalance be used to induce a SILC state? Our results show that the answer is yes.

In this letter, we map the existence and stability of the SILC state as the layer densities are unbalanced at constant $n_T$, spanning the entire range from balanced to completely imbalanced (all charge in one layer). We repeat this process for five densities (0.82 < $n_T$ < 1.8) × $10^{11}$ cm$^{-2}$, ranging from strongly interlayer coherent ($d/l_B$ = 1.26) to incoherent at balance ($d/l_B$ = 1.82). Our mapping of the stability of the SILC state versus $d/l_B$ and imbalance exhibits diverse behaviors, and we emphasize three key results. Firstly, we identify the point at which the bilayer system undergoes total charge transfer (all the charge in a single layer) and note that it is significantly enhanced by exchange-correlation effects. Secondly, we observe direct competition between the SILC state and the single-layer QH states in weakly coherent systems. Finally, our third and most significant result is that *a bilayer system that is incoherent at balance can develop SILC as the layers are imbalanced*.

The continuous evolution of the SILC state into the single-layer ν = 1 QH state with charge imbalance was established by Hamilton *et al.* [10]. Numerous authors have since studied this evolution in more detail, focusing on the stability of the SILC state, as measured by its energy gap $\Delta_{\nu=1}$ [7-9]. Recent theoretical calculations suggest that the SILC state is strengthened by imbalance [11,12]. In the Hartree-Fock approximation [12], $\Delta_{\nu=1}$ remains constant until the system is completely imbalanced, after which it increases linearly with increasing bias. In contrast, experimental studies have shown a continuous, parabolic increase in $\Delta_{\nu=1}$ with imbalance [7-9]. However, those studies were complicated either by the presence of a large tunneling gap ($\Delta_{SAS}$ = 6.8K [7] or 15K [8]) or restriction to a limited range of imbalance [9].

To overcome those limitations, we have used a bilayer hole sample consisting of a (311)A GaAs/AlGaAs heterostructure with two 15 nm wide GaAs quantum wells separated by a 2.5 nm wide AlAs barrier. The larger effective mass of holes, compared to electrons, suppresses interlayer tunneling, while *in-situ* back gates allow a wide gate range. To confirm the suppression of tunneling, Fig. 1 shows longitudinal ($\rho_{xx}$) and Hall ($\rho_{xy}$) resistivity measured as a function of $B$ at $d/l_B = 1.36$ ($n_T = 9.6 \times 10^{10}$ cm$^{-2}$) for the balanced bilayer system. The data in Fig. 1 shows only a single odd-ν QH state located at ν = 1, and a series of even-ν 'combined' QH states due to addition of independent single-layer QH states. The absence of QH states at odd-integer ν ≥ 3, as confirmed by Fig. 1 (inset), which focuses on the $\rho_{xx}$ oscillations at smaller $B$, indicates that interlayer tunneling is negligible ($\Delta_{SAS} < 0.1$K) and that the QH state at ν = 1 is entirely due to SILC. As an added advantage, bilayer hole systems allow measurements of devices with smaller $d$, and therefore larger $n_T$, where the effects of disorder are reduced [10].

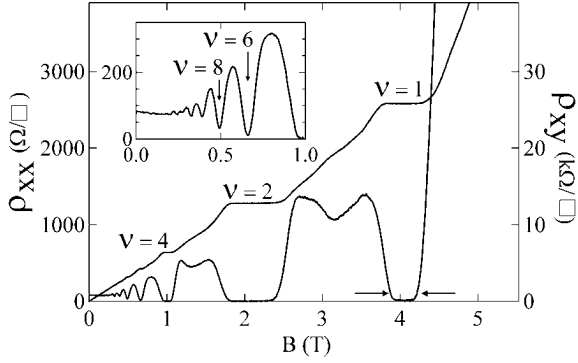

Fig. 1: Longitudinal $\rho_{xx}$ and Hall $\rho_{xy}$ resistivity traces obtained at balance ($n_t = n_b$) for $d/l_B = 1.36$ ($n_T = 9.7 \times 10^{10}$ cm$^{-2}$). Horizontal arrows indicate the width of the ν=1 QH minima $\Delta B$. Inset shows low-field $\rho_{xx}$ data, demonstrating the absence of odd-ν QH states.

Electrical measurements were performed using a four-terminal, low-frequency a.c. lock-in technique with an excitation current $I < 2$ nA at a lattice temperature of 55 mK. The two 2D hole layers are measured using a Hall bar geometry with AuBe ohmic contacts that contact both layers in parallel. The densities in the two layers are measured using $\rho_{xx}(B)$ data obtained at low $B$, and are controlled electrostatically using independently biased top and bottom gates above and below the top and bottom layers respectively. Imbalance at constant $n_T$ is achieved by adjusting the bias voltage on the top ($V_t$) and bottom ($V_b$) gates in opposite directions (i.e., one more positive and the other more negative) by amounts proportional to the respective gate-layer separation (see [13] for an extended discussion of this procedure).

To quantify charge imbalance, we introduce two parameters – the layer imbalance $m_z$ and the gate imbalance $N_z$. The layer imbalance $m_z = (n_t - n_b)/n_T$ is also known as the *z*-component of the pseudospin vector [4], and describes the fractional charge imbalance between the two layers. The layer imbalance takes values of $m_z = +1$, 0, and $-1$ when all holes are in the top layer, equally balanced, and in the bottom layer, respectively. The gate imbalance is the analogue of $m_z$ for the top and bottom gates, and describes the fraction of charge moved from one gate to another: $N_z = (\Delta N_t - \Delta N_b)/N_T$, where $N_T = N_t + N_b$ is the total surface charge on the gates, $\Delta N_{t,b} \propto V_{t,b}/d_{t,b}$ is the change in the top or bottom gate surface charge density from its value at balance, and $V_{t,b}$ is the bias applied to the top/bottom layer, assuming $V_{t,b} = 0$ V at balance. The distance $d_{t,b}$ is the gate-layer separation for the top/bottom layer. For a classical system, the gate and layer imbalances are equal, $N_z = m_z$. However, in our bilayer 2D hole systems, $|m_z| \geq |N_z|$, due to the strong negative compressibility caused by exchange-correlation effects [10].

Thermal activation studies are widely used to measure the energy gap of QH states. However, to undertake detailed thermal-activation studies over a wide range of imbalance for several different values of $d/l_B$ would require many months of measurements. Instead, we follow Sawada *et al.* [7], who have previously used the width in $B$ of the bilayer ν = 1 state as a comparative measure of its stability. Our approach is also based on studies of single-layer QH states [14], which have shown that the normalized magnetic field width $\Delta B/B$ of a QH state is directly representative of its energy gap. The normalization accounts for the dependence of $\Delta B$ on $n_T$, and we define $\Delta B$ to be the range in $B$ over which $\rho_{xx}$ is below a cutoff value $\rho_{xx}^c$, as indicated by the horizontal arrows around ν = 1 in Fig. 1. We note that over a reasonable range of $\rho_{xx}^c$, the precise value of $\rho_{xx}^c$ does not qualitatively affect our results. A non-zero (or zero) value for $\Delta B/B$ signifies the presence (or absence) of a QH state at ν = 1.

Figure 2 introduces a simplified phase diagram calculated for our bilayer QH system as a function of $N_z$ and $d/l_B$, based on comparing the ground-state energies of SILC states with those of two independent single-layer states [15]. The zero well width equivalents of the five values of $d/l_B$ used in the experiment are mapped onto the phase diagram in Fig. 2 as dashed horizontal lines, after accounting for finite well thickness [16]. Three distinct regions are evident in Fig. 2. The 'no SILC' phase (top center) corresponds to incoherent layers that do not exhibit SILC at $\nu = 1$. The 'SILC' phase (lower center) corresponds to the interlayer-coherent bilayer system with charge distributed across both layers, and exhibits a QH state at $\nu = 1$. The transition at $N_z = 0$ (no imbalance) from the 'SILC' to the 'no SILC' phase with increasing $d/l_B$ is well established [3,6]. Finally the $m_z = +1$ (far right) and $m_z = -1$ (far left) phases correspond to all of the charge being in the top or bottom layer, respectively. The continuous transition between the bilayer SILC phase and the single layer $m_z = \pm 1$ phases occurs for sufficiently large gate bias; the system minimizes its energy by moving all of the remaining charge into only one layer [4,11,12] and forming a single-layer $\nu = 1$ QH state. Note that this transition would occur at a significantly higher bias in the absence of exchange-correlation effects. The filled and open circles and the open square in Fig. 2 correspond directly to those shown in Fig. 3, and indicate the experimental observation of the respective transitions between phases.

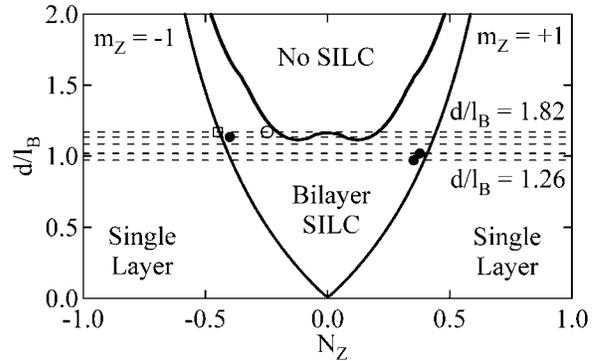

Fig. 2: Calculated ground-state phase diagram for a bilayer ν = 1 system, showing three distinct phases. Horizontal dashed lines correspond to the experimental data in Fig. 3, and are mappings of the measured $d/l_B$ corrected for finite well thickness (from top: 1.82, 1.66, 1.53, 1.35 and 1.26). Solid and hollow circles and square correspond to those shown in Fig. 3.

Figure 3 shows measurements of the stability $\Delta B/B$ of the bilayer $\nu = 1$ QH state obtained as a function of imbalance $N_z$ for five different values of $d/l_B$ ranging from 1.26 (top) to 1.82 (bottom). Our three key results are evident in this data; which we discuss in order of increasing $d/l_B$ and in conjunction with Fig. 2. Commencing with the lowest $d/l_B$ trace, we observe an approximately parabolic increase in $\Delta B/B$ with increasing $N_z$ for small imbalance $|N_z| < 0.15$, consistent with [7,9]. However, as $d/l_B$ is increased to 1.53 (middle trace in Fig. 3) the parabola acquires a flat bottom, as expected from Hartree-Fock theory [12]. It is unclear why this behavior has not been observed in previous experiments [7-9], and why the predicted behavior is only observed at larger $d/l_B$, since the Hartree-Fock calculations should improve as $d/l_B \to 0$. One possibility is that topological excitations [4] (not considered in [12]) cause deviations from the Hartree-Fock predictions at smaller $d/l_B$; another possibility is that disorder may play a role, since smaller $d/l_B$ requires low $n_T$, where the effects of disorder become more pronounced.

Our first key result is that for $|N_z| > 0.2$ all of the $d/l_B$ traces evolve similarly – $\Delta B/B$ rises sharply towards a clear peak and then decreases abruptly with increasing imbalance. These $\Delta B/B$ peaks are highlighted by the solid circles in the $d/l_B = 1.26$, 1.35 and 1.66 traces in Fig. 3, and are mapped onto the phase diagram in Fig. 2. The solid circles in Fig. 2 lie close to the calculated boundary between the SILC and $m_z = \pm 1$ phases. This indicates that the $\Delta B/B$ peaks correspond to the bilayer system undergoing complete charge transfer, transferring all holes into one layer in order to minimize energy. This charge transfer occurs at an $N_z$ that is significantly reduced due to exchange-correlation effects. We also note that that the $N_z$ where total charge transfer occurs decreases with decreasing density (i.e., reduced $d/l_B$), which is consistent with the shape of the SILC phase in Fig. 2.

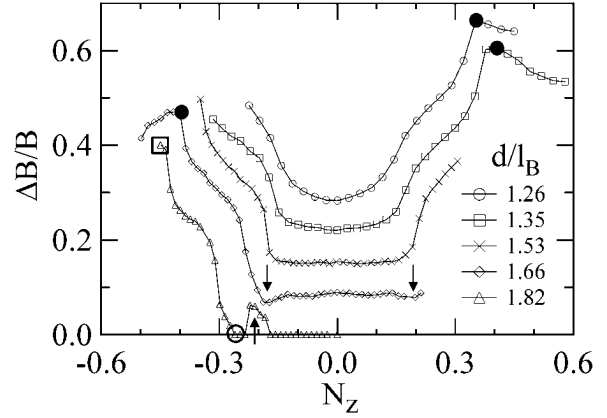

Fig. 3: Normalized width $\Delta B/B$ of the $\nu = 1$ $\rho_{xx}$ minima vs. layer imbalance $m_z$ for five $d/l_B$ values between 1.26 and 1.82. The arrows indicate the combined ⅓ + ⅔ state (see text) and the 3 solid and 1 hollow circles correspond to those shown in the phase diagram (Fig. 2). Consecutive traces are offset by $\Delta B/B = 0.04$ from each other for clarity.

Our second key result is evident in the weakly coherent system at $d/l_B = 1.66$. The downward pointing arrows in Fig. 3 indicate competition between the $\nu = 1$ SILC state and the combined states that originate from addition of independent single-layer fractional QH states. This is manifested as local minima in $\Delta B/B$ at $N_z$ values corresponding to combined states where intralayer correlations are strong (i.e., single-layer fractional QH states). The downward pointing arrows in Fig. 3 mark the point at which $\nu_t = ⅓$ and $\nu_b = ⅔$ (left), and $\nu_t = ⅔$ and $\nu_b = ⅓$ (right), for the $d/l_B = 1.66$ data. Both arrows coincide with local minima in the stability of the SILC state. To confirm our assertion that competition between the SILC state and the combined ⅓ + ⅔ state is indeed responsible for this effect, we track the evolution of the SILC $\nu = 1$ state and the four single-layer fractional QH states ($\nu_t = ⅓$, $\nu_b = ⅓$, $\nu_t = ⅔$ and $\nu_b = ⅔$) in Fig. 4. The grayscale plot shown in Fig. 4 consists of 36 separate traces of $\rho_{xx}$ (z-axis) vs. B (x-axis) obtained at different values of $N_z$ (y-axis). The dark regions in Fig. 4 correspond to $\rho_{xx}$ minima (i.e., QH states). We now identify the five relevant QH states in this plot. First, the thin diagonal dashed lines indicate the four single-layer fractional QH states: $\nu_t = ⅓$, $\nu_b = ⅓$, $\nu_t = ⅔$ and $\nu_b = ⅔$. The bilayer $\nu = 1$ state is the vertical dark band centered at $B \sim 6T$. The small light gray regions at $N_z = \pm 0.18$ in the bilayer $\nu = 1$ band correspond to the local minima in the $d/l_B = 1.66$ trace in Fig. 3, and are located where the bilayer $\nu = 1$ state is at its weakest. Most significantly, these gray regions coincide with the crossing points of the single-layer QH states where $\nu_t = ⅓$ and $\nu_b = ⅔$ or $\nu_t = ⅔$ and $\nu_b = ⅓$. This competition between the SILC state and the combined ⅓ + ⅔ state is not observed at smaller $d/l_B$ because the interlayer coherence becomes stronger and intralayer correlations become weaker as $d/l_B$ decreased (i.e., as $l_B$ increases).

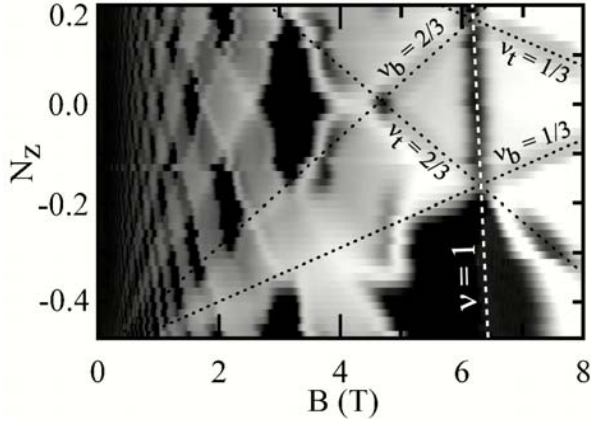

Fig. 4: Grayscale plot of $\rho_{xx}$ (z-axis) vs. $B$ (x-axis) and $N_z$ (y-axis). Dark regions correspond to QH states. The four diagonal dotted lines denotes single-layer fractional QH states. The vertical white dotted line corresponds to the bilayer $\nu = 1$ QH state.

Finally, and most importantly, we find that a bilayer system that is incoherent at balance can develop SILC with charge imbalance. Increasing $d/l_B$ to 1.82 (lowest trace in Fig. 3) weakens the interlayer coherence so much that the SILC state collapses ($\Delta B/B \rightarrow 0$) at $N_z = 0$, and the system behaves as two independent layers. As the layers are then imbalanced, proceeding left from the center in both Fig. 2 and 3, there is initially no QH state at $\nu = 1$ until at $N_z = -0.2$, where there is a well-defined peak in $\Delta B/B$ that indicates the formation of a combined $\nu = 1$ state with $\nu_t = \frac{1}{3}$ and $\nu_b = \frac{2}{3}$. This is verified using a grayscale plot similar to Fig. 4 (not shown). A small increase in imbalance beyond this peak destroys the combined state (i.e., $\Delta B/B$ returns to zero) as the intralayer correlations responsible for the combined state are no longer energetically favorable. At $N_z = -0.22$, indicated by the hollow circle at the bottom of Fig. 3, $\Delta B/B$ again becomes non-zero and begins to rise rapidly out to $N_z = -0.42$ for $d/l_B = 1.82$, indicated by the hollow square in Fig. 3. These two points in Fig. 3, the hollow circle and square, are mapped onto the phase diagram in Fig. 2 and suggest the origin of the rise in $\Delta B/B$. The circle lies immediately adjacent to the no SILC-SILC phase boundary, and the square is inside the SILC – $m_z$ = $\pm 1$ boundary, suggesting that the sharp rise in $\Delta B/B$ corresponds the SILC phase becoming the ground state of the bilayer system. It is interesting to note that the same steep rise in $\Delta B/B$ occurs for all values of $d/l_B$ once the imbalance exceeds some critical $N_z$, irrespective of whether a QH state exists at $\nu = 1$ for $N_z = 0$.

In conclusion, we used low-density, high-mobility bilayer hole systems with negligible tunneling to map the stability of the bilayer $\nu = 1$ QH state versus charge imbalance at constant total density for five values of $d/l_B$ ranging from strongly interlayer coherent ($d/l_B = 1.26$) to incoherent ($d/l_B = 1.82$) at balance. We observe three key results: (i) the significant enhancement of total charge transfer due to exchange-correlation effects, (ii) the presence of direct competition between the SILC state and single-layer fractional QH states in weakly coherent systems, and most significantly (iii) that bilayer systems that are incoherent at balance develop spontaneous interlayer coherence with imbalance. These results indicate a non-trivial relationship between charge imbalance and the stability of the SILC state. Experiments using devices with independent electrical contact to the layers would clarify the role of intralayer correlations in the effects reported here.

We acknowledge financial support from the Australian Research Council and NSF DMR-0206681 and EPS-0132626.